# Review on thermoelectric properties of transition metal dichalcogenides

I. Pallecchi [1], N. Manca [2], B. Patil [2], L. Pellegrino [2], D. Marré [1,3]

[1] CNR-SPIN, c/o Dipartimento di Fisica, via Dodecaneso 33, 16146 Genova, Italy
[2] CNR-SPIN, C.so Perrone 24, 16152 Genova, Italy
[3] Università di Genova, Dipartimento di Fisica, via Dodecaneso 33, 16146 Genova, Italy

**Abstract**
Transition metal dichalcogenides (TMDs) are considered an advantageous alternative to their celebrated two-dimensional (2D) van der Waals akin compound, graphene, for a number of applications, especially those requiring a gapped and possibly tunable band structure. Thermoelectricity is one of the application fields where TMDs could indeed outperform graphene, thanks to their lower thermal conductivity, large effective masses, valley degeneracy, varied and tunable transport properties, as well as sensitivity of their band structures and phonon spectra to confinement. Yet, despite promising theoretical predictions, thermoelectric properties of TMDs have not been extensively investigated so far and a clear assessment of TMDs as viable thermoelectric materials, based on experimental results, is still missing. In this paper, we review the experimental findings of literature on thermoelectric properties of TMDs, to sort out the countless combinations of chemical compositions, doping, off-stoichiometry and sample forms which could potentially result in optimized and possibly competitive thermoelectric properties. Based on the experimental data of literature, we simulate the performance of an all-TMD thermoelectric device for practical application as a micron sized cryocooler or power generator.

## 1. Introduction

Starting from the outstanding electrical, mechanical, thermal, and optical properties of graphene, research is recently veering towards other two-dimensional 2D van der Waals compounds, which could offer alternative benefits, expanded range of applications and improved performances in specific applications. Among these compounds, transition metal dichalcogenides (TMDs) can be considered the frontier of functional devices, thanks to their electronic, optical, spin, mechanical, topological and biological sensing properties [1,2,3,4,5], as well as to the intertwining of these properties [6] and their tunability by doping [7], strain [8,9], thickness [10] and electric fields [11,12]. Most notably, the gapped band structure of TMDs represents a favorable alternative to the gapless Dirac dispersion relationship of graphene for the fabrication of logic transistors [13,14], while large spin-orbit coupling in TMDs allows to induce spintronic functionalities in graphene by proximity in TMD/graphene heterostructures, such as spin-Hall effect and Rashba-Edelstein effect [15].

Thermal and thermoelectric properties of graphene and akin 2D materials have not been as much widely investigated so far. Graphene-based devices for thermal energy conversion can take advantage of thermoacoustic, thermoelectric and thermo-optical coupling, with the further perspective of potential scaling down to the nanoscale [16]. However, the record-high thermal conductivity of graphene, around 5000 $Wm^{-1}K^{-1}$ at room temperature [17], makes it intrinsically unsuitable for thermoelectric applications. On the other hand, the potential and perspectives of TMDs for thermoelectric applications is still to be realistically assessed. TMDs have much lower thermal conductivity than graphene, which can be attributed to their different band structure, scattering channels, and dispersion relationships calculated for the underlying phonon structure. Thermal conductivity $\kappa$ in TMDs is typically tens of $Wm^{-1}K^{-1}$ at room temperature along the in-plane direction [18], and lower along the out-of-plane direction, due to their layered structure and van der Waals gaps. Additionally, TMD samples can be prepared by liquid phase exfoliation, followed by restacking in the form of nanoflake assemblies, with tailored distribution of thickness and size, as well as with proper inter-flake connectivity [19]. In such samples, nanostructure can be engineered to increase scattering of phonons of different wavelengths, thus suppressing the lattice thermal conductivity. On the other hand, regarding electronic properties, TMD compounds exhibit a variety of transport behaviors that range from metallic, to semimetallic, to semiconducting. Accordingly, Seebeck coefficient S and electrical conductivity $\sigma$ may vary by orders of magnitude across different TMD compounds and even within the same compound, due to off-stoichiometry or field effect. Combination of high S and $\sigma$ yields enhanced power factor $S^2\sigma$, which is an indicator of the adequacy of electronic parameters for those thermoelectric applications where a heat source may provide unlimited power to maintain a thermal gradient, for example using recovery of waste heat. On the other hand, the further benefit of a low thermal conductivity $\kappa$ yields an enhanced figure of merit ZT= $S^2\sigma T/\kappa$, which measures the conversion efficiency of a thermoelectric material in terms of input and output power balance. Within this context, a huge worldwide effort on both theoretical and experimental grounds is needed to sort out the countless combinations of chemical compositions, doping, off-stoichiometry and sample forms (flakes, films, heterostructures, nanoflake assemblies, polycrystals) which could potentially allow to improve TMD performances for optimized thermoelectric applications.

Whereas the challenge of S and $\sigma$ measurements is usually related to the micrometric size of flakes, $\kappa$ measurements in 2D systems are even more difficult, as they require suspending nanomaterials, measuring the temperature distribution in nano/micro scale, and getting rid of thermal resistance of contacts. Furthermore, on the interpretative side, one must consider that thermal transport in 2D systems is significantly different from the case of three-dimensional counterparts, due to size and nonlinear effects, confinement of phonons and the influence of substrate phonons [20].

In this work, we provide an overview of literature theoretical predictions and experimental data about thermal and thermoelectric transport in TMDs.

## 2. Remarkable thermoelectric properties of some TMD-related materials

The potential interest in TMDs as thermoelectric materials arises from the observation of interesting behavior of thermoelectric properties in other materials akin to TMDs. Among these, besides graphene, also SnSe and $ZrTe_5$ are worth mentioning, the former being an outstanding anisotropic thermoelectric, the latter exhibiting thermoelectric behavior related to its topological character.

### 2.1. Graphene

Graphene, due to the presence of light atoms and strong in-plane bondings, exhibits $\kappa$ values as high as 5000 $Wm^{-1}K^{-1}$ [17], while its Seebeck coefficient is reduced by opposite contributions of electrons and holes. Both these characteristics make graphene unsuitable for thermoelectric applications. However its phonon and electronic properties can be manipulated and tailored for specific uses, for example by inducing a bandgap with lateral confinement or with nanomeshes of holes. Theoretical predictions on graphene nanoribbons indicate that S values up to 4mV/K [21] and ZT values exceeding 3 [22,23] could be obtained.

On the experimental side, enhanced thermoelectric performance was demonstrated on as-grown suspended graphene nanoribbons [24], where a large reduction of lattice thermal conductivity was obtained thanks to size-controlled ballistic phonon transport. Simultaneously, enhanced S (-127 $\mu$V/K at 220K) as compared to bulk value was achieved thanks to bandgap opening, thus yielding ZT values up to ~0.12.

At high temperature, S is substantially larger than the prediction of the Mott relation, approaching the hydrodynamic limit due to strong inelastic scattering among the charge carriers. Close to room temperature the inelastic carrier–optical-phonon scattering becomes more significant and limits S below the hydrodynamic prediction [25]. Seebeck coefficient around ~80 $\mu$V/K was measured in graphene at 300 K by several groups [25,26,27].

In the quantum Hall regime at a high magnetic field, quantized Seebeck and Nernst signals were observed [26,27,28]. Divergent S and large Nernst signal were observed for Fermi level crossing the Dirac point [28].

Graphene properties can be also boosted in composites. Room-temperature thermoelectric power factor 0.0187 $mW\, m^{-1}K^{-2}$ was obtained in graphene-based organic nanocomposites deposited by ink-jet printing on large flexible substrates for flexible thermoelectric device applications [29].

### 2.2. SnSe

The recent discovery of giant figure of merit, ZT up to 2.6 at 923 K, in p-type SnSe monochalcogenide single crystals was considered a breakthrough in the search for new thermoelectric materials [30,31]. In this anisotropic compound, the low thermal conductivity is thought to originate from high anharmonicity of the chemical bonds [32], while high power factor is thought to come from its high electrical conductivity and

Seebeck coefficient, enabled by multiple electronic valence bands [31,33] and quasi-2D band dispersion [33]. SnSe polycrystals have generally higher $\kappa$, lower $\sigma$ and thus lower ZT than single crystals, resulting in lower ZT values, for example ZT=0.5 at 820K [34] and ZT=0.33 at 350 K in 1 at. % Na-doped samples [35]. In SnSe polycrystals, both p-type and n-type thermoelectrics, optimized for high temperature operation, were obtained by chemical doping, namely p-type indium doped SnTe polycrystals with ZT≈1.1 at 873K [36], p-type silver doped SnSe polycrystals with ZT≈0.6 at 900K [37], p-type silver doped $Ag_{0.01}Sn_{0.99}Se_{1-x}S_x$ polycrystals with $\kappa$ as low as 0.11 W m$^{-1}$K$^{-1}$ and ZT=1.67 at 825K for x=0.15 [38], n-type iodine doped SnSe polycrystals with ZT≈0.8 at 773K [39], n-type iodine doped Sn(Se,S) polycrystals with ZT≈1.0 at 773K [39].
Nanostructuring was also attempted to improve thermoelectric performance of SnSe films [40] and nanoflakes [41,42]. In ref. [41], $SnSe_{1-x}S_x$ nanosheets were prepared from bulk ingots via lithium intercalation, to ease the exfoliation process. Lowering of $\kappa$ and enhancement of ZT (ZT=0.12 at 310K) was thus obtained by combining three different strategies: (i) substitution with isoelectric atoms Se and S (point defect phonon scatterers), (ii) nanostructuring via exfoliation of nanosheets from the bulk material, (iii) chemical transformation of the material into a porous structure. The latter porous structure of $SnSe_{1-x}S_x$ nanosheets was obtained through solution-phase chemical transformation using tartaric acid (t-acid) and $O_2$. In ref. [42], low thermal conductivity ($\kappa$=0.08 W K$^{-1}$m$^{-1}$) was obtained in thermally evaporated porous networks of SnSe nanosheets, thanks to a combination of porosity and thermal resistance at the grain boundaries. However, in these systems with preferential (111) orientation the measured power factor was lower than in SnSe polycrystals and ZT peaked at 0.05 at 500K.

### 2.3. ZrTe$_5$
Thermoelectric transport in topological materials is a novel topic of remarkable fundamental and applicative relevance. Zirconium pentatelluride is a three-dimensional Dirac semimetal, whose thermoelectric behavior reflects its topological phase.
In ref. [43], at low temperature, a step-like profile near zero field of Nernst effect versus magnetic field was measured, which is a signature of the anomalous Nernst effect (ANE) resulting from the Berry curvature. Seebeck and Nernst signals presented signatures of entering the extreme quantum limit (one single Landau Level occupied) with increasing magnetic field. With further increasing magnetic field, the system was turned gapless, leading to a large dip for the Seebeck signal and a sign change for the Nernst signal.
The potential applicative interest of ZrTe$_5$ thermoelectric behavior lies in the significant increase of Seebeck and Nernst signals with magnetic field in the extreme quantum limit [44]. No less interesting are the fundamental aspects, related to the Berry curvature and to the thermoelectric Hall conductivity $\alpha_{xy}=S_{xy}/\rho$
The latter, in particular, was observed to acquire a universal, quantized value deep in the extreme quantum limit, which is independent of magnetic field or carrier concentration, and linear in T, indicating the presence of three-dimensional Dirac or Weyl electrons [44].

### 3. Theoretical predictions on thermoelectric properties of TMDs
A number of theoretical studies indicate that TMDs could be superior thermoelectric materials and their thermoelectric behavior could be further improved by confinement. In p-type and n-type $MoS_2$ and $WSe_2$, the figure of merit was predicted to be close to unity and to vary with the number of atomic layers, being maximum for one or two monolayers (MLs) [45,46]. The latter dependence on confinement was obtained from ab initio calculations of phonon dispersion, indicating a lowering of thermal conductivity with decreasing number of MLs. Similarly, as a consequence of the effect of quantum confinement on the phonon spectrum and electronic structure, a negative correlation between power factor and $\kappa$ was predicted in $SnS_2$ [47]. It was calculated that by diminishing the thickness of $SnS_2$ nanosheet to about 3 MLs, S and $\sigma$ along the in-plane directions would simultaneously increase, whereas $\kappa$ would decrease, achieving a high ZT value of 1.87 at 800 K, at optimal carrier concentration. In ref. [48], tuning of the band structure, and thus of thermoelectric properties, by varying the thickness was calculated in one to four MLs thick $MoS_2$, $MoSe_2$, $WS_2$, and $WSe_2$. In all these cases, the maximum ZT coincided with the greatest near-degeneracy within $k_BT$ ($k_B$ Boltzmann constant) of the band edge that resulted in the sharpest turn-on of the density of modes. The thickness at which this maximum occurs was, in general, not a single ML. Large thermoelectric power factor was predicted for TMDs with 1T'' phase [49]. In particular, high mobility and weak electron-phonon coupling in 1T''

MoSe$_2$ yielded large power factors of ~$6.0\times10^{-3}$ W/mK$^2$ in the temperature range from 100 to 500 K. In ref. [50], tensile strain was predicted to enhance the figure of merit in ZrS$_2$, ZrSe$_2$ and ZrTe$_2$ monolayers by simultaneously reducing the lattice thermal conductivity and increasing the power factor, with maximum ZT in correspondence of the largest bandgap. The influence of spin-orbit interaction on thermoelectric properties was calculated in MoS$_2$ nanoribbons with armchair and zigzag edges [51]. In ref. [52], calculations of the Seebeck coefficient of TiS$_2$ nanosheets indicated an increase with decreasing thickness, with S up to 40% larger in the ML case with respect to the bulk TiS$_2$. In parallel, the acoustic phonon band in the ML was calculated to be flatter than that of the bulk, which should be beneficial for a low lattice thermal conductivity. Thermoelectric properties of n-type SnSe$_2$ were predicted to be highly anisotropic and significantly better than those of p-type SnSe$_2$ [53]. Figures of merit along in-plane and out-of-plane directions ZT$_a$ =2.95 and ZT$_c$ = 0.68 at optimal doping (n = $10^{20}$ cm$^{-3}$) and 800 K were calculated, so that in polycrystalline samples ZT=0.75 at 300K and ZT= 3.5 are expected. In cubic metastable ZnSe$_2$ very low κ~0.69 W/(m·K) at room temperature was predicted from calculation of phonon spectrum, which, combined with calculated power factor values S$^2$σ≈1.5 mW m$^{-1}$K$^{-2}$ for both p-type and n-type doping, resulted in ZT≈0.5 [54]. A number of TMDs featured among the binary semiconducting chalcogenide compounds screened by high-throughput computational method, as expected to have high power factor and low lattice thermal conductivity [55].

Prediction on TMD heterostructures are also promising. In MoS$_2$/WS$_2$ hybrid nanoribbons [56], the MoS$_2$/WS$_2$ interface was predicted to reduce lattice thermal conductivity more than the electron transport, yielding ZT~5.5 at 600 K. In ref. [57] high n-type power factor related to high electronic band degeneracy and glassy phonon transport determined by interface scattering were predicted in superlattice-ML ZrSe$_2$/HfSe$_2$. ZT ≈5.3 at 300 K in the n-type superlattice and ZT≈3.2 at 300 K in the p-type superlattice were calculated. In ref. [58] figures of merit ZT~1.1 and ZT~0.3 were predicted in ultrathin BP/MoS$_2$ bilayers for p-type and n-type doping, respectively, at 300K and even gigantic ZT>4 values at 800K.

In summary, the large effective masses and valley degeneracies, as well as effect of confinement on the band structure, on scattering rate and on phonon spectrum are promising predictions for considering TMDs as thermoelectric materials.

**4. Experimental results on TMD flakes, films and polycrystals**

Thermoelectric properties measured in TMD single crystalline flakes and films reflect intrinsic transport properties, related to the band structure of the materials. Hence, flake samples are a perfect playground to investigate the thermoelectric potential of these compounds. Highly dense and highly oriented polycrystals may exhibit very similar electric and thermoelectric transport properties to their single crystalline counterparts. On the other hand, interfaces and random orientation of grains in polycrystals may offer a further opportunity to tailor thermoelectric and thermal transport, with the additional benefit of easy handling and mechanical robustness for processing and practical application.

Thermoelectric properties in TMDs have been measured mostly in MoS$_2$, however several works also explore other TMD compounds such as SnS$_2$, TiS$_2$, WSe$_2$, PtSe$_2$ and 1T'-MoTe$_2$.

**4.1. MoS$_2$**

The most remarkable result within the existing literature on thermoelectric properties in TMDs is the finding of a power factor PF as large as 8.5 mW m$^{-1}$K$^{-2}$ at 300K in two-ML MoS$_2$ flakes [59] (see Fig. 1, upper left box), larger than the value of the commercial thermoelectrics Bi$_2$Te$_3$ and PbTe. This value was obtained in correspondence of a field effect induced carrier concentration of n$_{2D}$~1.06x10$^{13}$ cm$^{-2}$, in the regime of n-type degenerate semiconductor. Indeed, the boost of the Seebeck coefficient was attributed to the confinement-dependent large effective mass and valley degeneracy obtained as the Fermi level approached the bottom of the conduction band. Moreover, the power factor versus gate voltage curves indicated that the power factor may further increase for even larger carrier concentrations in one- to three-MLs thick samples. The authors also suggested that a further increase of the Seebeck coefficient could be obtained by engineering the dominating scattering mechanism (e.g. by charged impurities rather than by phonons), which determines the energy dependence of the scattering rate and enters the Mott formula for S. Very similar results were obtained in ref. [60], where thickness dependent and gate tunable electrical and thermoelectric properties of few-layer $MoS_2$ flakes were measured (see Fig. 1, bottom box). The conductance was shown to increase significantly with decreasing thickness, thus yielding a power factor up to 3.0 mW m$^{-1}$K$^{-2}$ at 300K in the two-ML sample, in the n-type accumulation regime.

Thermal conductivity measurements of $MoS_2$ carried out by different methods, among which Raman spectroscopy and thermal bridge, indicate room temperature values in the range 35-55 Wm$^{-1}$K$^{-1}$ [18]. In ref. [61], κ was extracted from the measurement of thermal time constants in suspended single-layer $MoS_2$ drum shaped flakes, using thermomechanical response to a high-frequency modulated laser. Here, κ values in the range 10-30 Wm$^{-1}$K$^{-1}$ were obtained, with a considerable spread between devices, likely caused by microscopic defects that have a large impact on phonon scattering. On the other hand, by proper sample manipulation, phonon scattering can be tailored and significant suppression of thermal conductivity can be obtained. In ref. [62] defects were induced in suspended few-layer $MoS_2$ flakes by increasing exposure time to plasma oxygen. Increasing disorder from the crystalline to the amorphous limit was measured by Transmission Electron Microscopy (TEM) and the corresponding κ varied from 30-35 Wm$^{-1}$K$^{-1}$ to 1-3 Wm$^{-1}$K$^{-1}$, respectively.

Thermal conductivity in TMD is highly anisotropic, indeed $MoS_2$, and in general TMDs, have very low cross-plane thermal conductivity. However, along the cross-plane direction power factors are small, and hence also the thermoelectric figure of merit ZT is small along this direction. On the other hand, power factors are large in the in-plane direction, yet large in-plane thermal conductivity in single crystals or ML films yields again small ZT (<0.05) also in the in-plane direction. This no-win situation was addressed in ref. [63], where layered $MoS_2$ polycrystals with high degree of orientation and with $MoO_2$ nanoinclusions were prepared by solid state reaction, with subsequent densification by Spark Plasma Sintering (SPS) at high temperature under a uniaxial pressure (50 Mpa). In these samples, the n-type oxygen doping of $MoS_2$ and the presence of metallic $MoO_2$, having with much larger (5 orders of magnitude) conductivity than $MoS_2$, simultaneously enhanced the electrical conductivity σ and the Seebeck coefficient S, especially in the in-plane direction. At the same time, the thermal conductivity κ in the cross-plane direction was reduced due to the formation of secondary $MoO_2$ phase. Indeed, due to lattice mismatch and difference in lattice vibrations in $MoO_2$ and $MoS_2$, interface scattering of phonons decreased thermal conductivity, especially in the cross-plane direction. In quantitative terms, a ZT of about 0.14 was obtained at 760K along the cross-plane direction, which means an improvement by 50 times in layered $MoO_2$/$MoS_2$ as compared to undoped $MoS_2$. Also the power factor was improved by 30 times and κ was decreased almost twice at room temperature in layered $MoO_2$/$MoS_2$. Based on a similar approach, p-type $MoS_2$ samples with $VMo_2S_4$ nanoinclusions exhibited enhanced thermoelectric performances [64]. Indeed, p-type Vanadium doping in the Mo site enhanced electrical conductivity by increasing both carrier concentration and mobility and simultaneously the formation of secondary $VMo_2S_4$ phase was effective in scattering phonons at the interfaces and in decreasing κ. In such samples, ZT≈0.18 at 1000K was obtained.

**4.2. $SnS_2$**

Thermal and thermoelectric transport was measured in tin disulphide $SnS_2$ nanoflakes, using a focused laser to create a temperature gradient [65] (see Fig. 2). Remarkably, an inverse correlation between electrical conductivity σ and thermal conductivity κ with varying thickness was obtained, namely as the thickness decreased, σ increased whereas κ decreased. This result suggested opportunities of improvement. For a 16-nm-thick sample, S ≈35 mV/K and κ ≈3.45 Wm$^{-1}$K$^{-1}$ were obtained at room temperature, resulting in a figure of merit value ZT≈0.13, that is 1000 times larger than the bulk crystal value.

**4.3. $SnSe_2$**

Thermoelectric properties of tin diselenide $SnSe_2$ have been object of several investigations, as this compound is chemically akin and similarly layered as its monochalcogenide counterpart SnSe, whose figure of merit ZT up to 2.6 at 923 K has risen worldwide interest [30,31].

$SnSe_2$ samples were fabricated in the form of hot (400°C) pressed pellets of nanosheets [66]. In such samples, a thermal conductivity as low as κ ≈0.67 Wm$^{-1}$K$^{-1}$ at room temperature was measured, owing to efficient phonon scattering by both the interfaces between the layers of $SnSe_2$ and the nanoscale grains in the nanosheets samples. Regarding electronic properties, both undoped and chlorine doped $SnS_2$ pellets resulted to be n-type. Most interesting, both |S| and σ were found to simultaneously increase with increasing temperature, at opposite with their inverse mutual correlation usually found and expected on the basis of the opposite dependence of these quantities on carrier concentration, according to Mott and Drude relationships, respectively. Despite the power factor values were not particularly large, $S^2\sigma \approx 0.053$ mW m$^{-1}$K$^{-2}$ at room temperature and $S^2\sigma \approx 0.146$ mW m$^{-1}$K$^{-2}$ at 610K in the chlorine doped sample, promising values of the figure of merit were obtained, ≈0.22 and ≈0.15 for the chlorine doped and undoped SnSe$_2$ at 610 K, respectively. SnSe$_2$ nanosheets pellets were also prepared by a SPS process, which allowed to obtain preferential crystalline orientation of (00l) facets [67]. The usual behavior that σ increases with doping and |S| decreases with doping was observed. Enhanced power factors along the in-plane direction were achieved via simultaneously introducing Se deficiency and chlorine doping, while efficient scattering of phonons of different wavelengths was determined by nanostructural and microstructural features. Along the in-plane direction, at room temperature $S^2\sigma \approx 0.8$ mW m$^{-1}$K$^{-2}$ and $\kappa \approx 2$ Wm$^{-1}$K$^{-1}$ resulted in ZT≈0.15, while the lower thermal conductivity at higher temperature determined a remarkable ZT≈0.63 at 673K. In similar highly oriented and dense SnSe$_2$ prepared by SPS, the beneficial effect of simultaneous copper intercalation and barium substitution on the thermoelectric performance was demonstrated by experiment and theory [68], remarkably in a single band context, and contrary to the general approach to improve thermoelectric performance. Cu ions are intercalated into wide van der Waals gaps and act as donors, while Ba$^{2-}$ substituted in the anionic sites of SnSe$_2$ planes are strong electron acceptors and facilitate the charge transfer from Cu$^+$ to the SnSe$_2$ layers. The result is an improvement of charge-carrier mobility and concentration, without degradation of the Seebeck coefficient, which improves the thermoelectric performance. SnCu$_{0.005}$Se$_{1.98}$Br$_{0.02}$ samples showed carrier concentration and mobility higher values than those of the control samples of SnSe$_2$, SnCu$_{0.005}$Se$_2$, and SnSe$_{1.98}$Br$_{0.02}$ over the entire temperature range. Power factor ≈1.2 mW m$^{-1}$K$^{-2}$, almost temperature-independent from 300 to 773 K was measured, which was the highest reported for all polycrystalline SnSe$_2$ - and SnSe-based materials. Figure of merit ZT ≈0.67 at 773K and ZT ≈0.10 at 300K were also measured. In Fig. 3, thermoelectric properties of highly oriented and dense SnSe$_2$ polycrystals are collected and compared.

The effect of microstructure on thermoelectric properties was investigated in SnSe$_2$ films deposited by physical vapor deposition and annealing in Se partial pressure [69]. In these films, structural disorder was introduced by random in-plane orientation among successive layers, while maintaining high degree of c-axis orientation. After Se annealing, which mitigated Se vacancies and promoted SnSe-to-SnSe$_2$ phase transition, cross-plane Seebeck coefficients up to -630 μV/K and power factors up to 2x10$^{-4}$ mW m$^{-1}$K$^{-2}$ at room temperature were measured.

### 4.4. PtSe$_2$

PtSe$_2$ is a less studied TMD, yet promising results for thermoelectricity were found. In PtSe$_2$ nanosheets, a thermopower enhancement up to values higher than 1 mV/K, more than 50 times than that of the bulk, was obtained with decreasing thickness down to 2 ML [70]. Indeed, with decreasing thickness, there is a bandgap opening and a transition from semimetallic to semiconducting character. In the semiconducting regime, transport properties become highly tunable, and thus the thermoelectric properties can be effectively optimized by field effect gating and extrinsic doping.

### 4.5. TiS$_2$

Noteworthy thermoelectric potential is found in semimetallic titanium disulphide TiS$_2$, whose transport properties, including Seebeck coefficient [71], are strongly affected by possible large off-stoichiometry and intercalation. In TiS$_2$ crystals, exhibiting semimetallic and highly anisotropic behavior, S≈-251 μV K$^{-1}$, power factor 0.371 mW m$^{-1}$K$^{-2}$ and $\kappa \approx 6.8$ Wm$^{-1}$K$^{-1}$ were obtained at 300K along the in-plane direction [72], resulting in ZT≈0.16. In this work, S and σ values were explained in terms of large density of states just above the Fermi energy and inter-valley scattering. TiS$_2$ can be very easily intercalated, for example with copper [73] and neodymium [74]. Intercalated TiS$_2$ were prepared by SPS, which allowed obtaining dense samples with high degree of crystalline orientation. Either Cu or Nd intercalation into TiS$_2$ leaded to substantial decrease in both electrical resistivity (controlling n-type charge carriers) and lattice thermal conductivity (introducing disorder) as compared to pristine $TiS_2$. For $Cu_{0.02}TiS_2$, ZT≈0.45 at 800 K and power factor 1.7 mW m$^{-1}$ K$^{-2}$ at 325 K were obtained [73], while for $Nd_{0.025}TiS_2$ ZT≈0.027 at 300 K was obtained [74].

**4.6. WSe$_2$**

A record low thermal conductivity κ≈0.05 W m$^{-1}$K$^{-1}$ at 300K along the cross-plane direction was obtained in disordered, layered WSe$_2$ crystals [75]. This value is 30 times smaller than the c -axis thermal conductivity of single-crystal WSe$_2$ and 6 times smaller than the predicted minimum thermal conductivity for WSe$_2$. This result was obtained in films with random stacking of WSe$_2$ layers, deposited by modulated elemental reactants method. All the WSe$_2$ layers were perfectly aligned along the c axis but had random relative in-plane orientation, according to the same concept used in ref. [69] for disordered SnSe$_2$ films, and this strategy turned out effective in the localization of thermal vibrations.

**4.7. 1T'-MoTe$_2$**

1T'-MoTe$_2$ is a Weyl semimetal, undergoing a monoclinic (non polar) to orthorombic (polar) transition at 250K. The transition temperature can be suppressed by applied pressure [76] or doping [77] and in both cases the low temperature Seebeck coefficient is dramatically enhanced close to the polar structural instability, because of the enhancement of electron–phonon scattering by the softening of phonon modes. In 1T'-MoTe$_2$ single crystal under applied pressure, the concomitant enhancement of S and σ yielded giant power factor of 30 mW m$^{-1}$ K$^{-2}$ at 25K, while in Nb doped 1T'-MoTe$_2$ polycrystals, the lower σ determined a maximum power factor 3.6 mW m$^{-1}$ K$^{-2}$ at 50K for 8% doping.

**4.8. Experimental results on TMD heterostructures and alloys**

Improvement of thermoelectric performances was pursued in TMD superlattices, taking advantage of the role of interfaces as phonon scatterers, as well as of the combination of materials with different structural and transport properties. In SnSe$_2$/MoSe$_2$ lattice-mismatched heterostructures [78], low temperature synthesis enhanced the turbostratic disorder (random rotation or translation between adjacent basal planes) induced by the lattice mismatch and was responsible for the ultralow thermal conductivity, down to 0.04 Wm$^{-1}$K$^{-1}$ in the 8 unit cell thick superlattice. In $(MS)_{1+x}(TiS_2)_2$ (M = Pb, Bi, Sn) superlattices [79], with alternating rock-salt-type MS layers and paired trigonal anti-prismatic TiS$_2$ layers, lattice symmetry mismatch and weak interlayer bonding were responsible for the low κ, while maintaining metallic behavior thanks to charge transfer across MS and TiS$_2$ layers. Power factors up to ~1 mW m$^{-1}$K$^{-2}$ at 500K were measured in $(SnS)_{1.2}(TiS_2)_2$ and ZT≈0.28-0.37 at 700K in all $(MS)_{1+x}(TiS_2)_2$ compositions. With a wider selection of compounds for TiS$_2$-based superlattices, $(MX)_{1+x}(TX_2)_n$ (where M = Pb, Bi, Sn, Sb or RE; T=Ti, V, Cr, Nb or Ta; X= S, Se; n=1, 2 or 3), both p-type and n-type carriers can be obtained [80]. In the $(SnS)_{1.2}(TiS_2)_2$ member of the series, the TiS$_2$ layer was found to provide the electron pathway to electric and thermoelectric transport, while the intercalated SnS layer suppressed the phonon transport, yielding a lowered in-plane thermal conductivity κ≈2.4 Wm$^{-1}$K$^{-1}$ and ZT≈0.37 at 300K. Analogous misfit-layered compounds were prepared by either hot [81] or cold pressing [82] in the form of highly textured polycrystalline alloys $(SnS)_{1.2}(TiS_2)_2$. By chemical substitution in the Sn site with Sb [81] and in the Ti site with Co and Cu [82], significant improvements of thermoelectric performances were observed, thanks to multifold effects. In particular, in-plane power factors were enhanced by texturing and optimized charge density, reaching ~0.8 mW m$^{-1}$K$^{-2}$ at room temperature and ~0.9 mW m$^{-1}$K$^{-2}$ at 570K in samples 4% Sb doped in the Sn site [81] and similar values ~0.6 mW m$^{-1}$K$^{-2}$ at room temperature and ~0.7 mW m$^{-1}$K$^{-2}$ at 720K in samples 2% Cu and Co doped in the Ti site [82]. Simultaneously, thermal conductivity was decreased thanks to mass enhancements, point defect scattering and weakened sound velocity induced to chemical substitution, and by interlayer scattering. Similar figure of merit values ZT≈0.11 at room temperature and ZT≈0.40 at 720K were measured in sample substituted either in the Sn site [81] or in the Ti site [82]. In misfit-layered $(BiS)_{1.2}(TiS_2)_2$ alloys, doped with transition metal elements on the Ti site, only a minor improvement of ZT was obtained by Cr substitution with respect to the pristine $(BiS)_{1.2}(TiS_2)_2$ alloys [83].

TiS$_2$ was also combined with organic materials in hybrid superlattices of alternating TiS$_2$ MLs and organic cations, fabricated by chemical vapor transport and subsequent electrochemical intercalation, for flexible thermoelectric materials [84]. In these systems, n-type power factor ≈0.45 mW m$^{-1}$K$^{-2}$ (45 μW cm$^{-1}$K$^{-1}$), in-plane thermal conductivity κ≈0.69 W m$^{-1}$K$^{-1}$ and ZT≈0.2 were measured at room temperature, even improved at higher temperature, ZT ≈ 0.28 at 373 K.

## 5. Experimental results on TMD nanoflake assemblies and TMD-based composites prepared by liquid phase exfoliation

Thanks to weak van der Waals bondings, TMD samples can be prepared in the form of nanoflake assemblies by liquid phase exfoliation and restacking by drop-casting or ink-jet printing. Nanoflake assemblies have peculiar transport properties as compared to flakes and films, mostly determined by the degree of interflake connectivity and degree of interflake barrier transparency to electron and phonon transport, as well as by the presence of unintentional defects and impurities associated to the preparation process. In addition, in chemical exfoliation, lithium intercalation tends to convert the TMD phase structure, favoring a phase transition from the semiconducting (trigonal prismatic) 2H to the (octahedral) 1T metallic phase, the latter having much larger conductivity and better thermoelectric properties. Nanoflake assemblies may be a viable route to large-scale production and may be advantageous is terms of tradeoff between quality/performance and cost/simplicity of fabrication process. In general, both electronic and thermal conductivity of nanoflake assemblies are much smaller than those of single crystalline samples, but the resulting thermoelectric figure of merit may be notable.

Assemblies of MoS$_2$ nanosheets were fabricated by liquid phase exfoliation, deposition, pressing against the PET substrates, and thermal annealing in vacuum to remove solvents [85]. The resulting samples were mixed phase, namely 2H-MoS$_2$ and 1T-MoS$_2$. The former, the more stable one, exhibited n-type semiconducting behavior, while the latter exhibited semimetallic and predominantly p-type behavior. A 10$^4$ times higher electrical conductivity and 100 times larger power factor were measured in the 1T-phase as compared to the 2H phase. Thermal treatment at 50°C maximized the fraction of 1T phase, while annealing at higher temperature increased the relative fraction of 2H phase, to the detriment of thermoelectric properties. In 1T-MoS$_2$ nanosheets S≈85.6 μV/K and σ ≈9978.3 Ω$^{-1}$ m$^{-1}$ resulted in a power factor 0.0731 mW m$^{-1}$K$^{-2}$ at room temperature. Unfortunately, despite the interesting control of thermoelectric properties by fabrication parameters, these systems suffered of a dramatic detrimental effect of exposure to environmental humidity.

Decoration of TMD nanoflake assemblies with metallic nanoparticles can be obtained by adding solutions of salts of the chosen metal into liquid dispersion of chemically exfoliated TMD nanosheets and by subsequent vacuum filtering. A "chemical welding" with Al$^{3+}$ ions was realized in n-type semimetallic TiS$_2$ nanosheets assemblies on flexible substrates [86]. Specifically, nanosheets deposited from liquid phase exfoliation, resulted to be bridged with multivalent cationic metal Al$^{3+}$, which cross-linked the nearby sheets. In Al$^{3+}$-welded assemblies a simultaneous increase of S and σ as compared to Al$^{3+}$-free assemblies was observed and possibly explained as an electrostatic effect of Al$^{3+}$ ions on TiS$_2$ bands. Consequently, a considerable power factor 0.217 mW m$^{-1}$K$^{-2}$ was measured on Al:[TiS$_2$ nanosheets] assemblies, which is very high a value for nanosheets assemblies. Unfortunately, dramatic detrimental effect of exposure to environment of both Al:[TiS$_2$ nanosheets] and 1T TiS$_2$ nanosheets was observed. From closer analysis, these Al:[TiS$_2$ nanosheets] were actually found to be Al:[H$_{0.5}$TiS$_2$ nanosheets]. In 1T-MoS$_2$ nanosheet assemblies decorated with copper cations [87], the Cu cations tended to distribute in the defect areas at the edges of MoS$_2$ nanosheets, rather than intercalating in interlayer spacing, due to more negative surface potential at the nanosheet edges than at the nanoshhet faces. As a consequence of charge transfer between MoS$_2$ and Cu, as well as of energy filtering mechanism [88] at the nanosheet boundaries, the MoS$_2$-Cu system exhibited 15 times larger mobility, 123 times lower carrier density, 2.6 times higher Seebeck coefficient up to ∼160μV/K and enhanced thermoelectric power factor up to 0.0235 mW m$^{-1}$ K$^{-1}$ as compared to the pure MoS$_2$ system. Also environmental stability of 1T-MoS$_2$ in 40-50% humidity improved by Cu cation addition. By adding a AgNO$_3$ water solution into an aqueous dispersion of chemically exfoliated 1T-MoS$_2$ nanosheets spontaneous Ag nanoparticles decoration was obtained [89]. Electrical conductivity first increased with increasing AgNO$_3$ molar ratio and nearly doubled at 0.5% mol, due to charge transfer and energy filtering at the nanosheet boundaries between MoS$_2$ and Ag nanoparticles, and then it decreased at higher molar ratios, due to aggregation of particles and consequent increase in the overall interface area. Carrier density peaked at 0.5%mol, where it was increased by a factor larger than 30 with respect to the pure MoS$_2$. On the other hand, Seebeck coefficient monotonically increased with AgNO$_3$ molar ratio, reaching ~140 µV/K and the resulting power factor was maximum at 7%, ~0.030 mW m$^{-1}$ K$^{-1}$.

Composites of reduced graphene oxide (rGO) and 1T-MoS$_2$ and 1T-WS$_2$ were prepared by liquid phase exfoliation and vacuum filtration [90]. At low rGO weight ratio, charge transfer at the rGO-MoS$_2$ and rGO-WS$_2$ heterojunctions improved electrical conductivity as compared to rGO-free samples, without a significant reduction of the Seebeck coefficient. The power factors peaked at 2-3% wt of rGO content, reaching values around ~0.015-0.017 mW m$^{-1}$ K$^{-1}$, which are nearly twice the values of respective pure MoS$_2$ and WS$_2$ systems. Thermal conductivities of rGO-MoS$_2$ and rGO-WS$_2$ as low as ~0.2 W m$^{-1}$ K$^{-1}$ were measured for 2-3% wt of rGO, yielding noteworthy ZT values 0.022-0.025.

Hybrid composites of TMD nanoflakes and organic materials offer remarkable opportunities. Poly(3,4-ethylenedioxythiophene):poly(4-styrenesulfonate) (PEDOT:PSS) is an organic polymer characterized by mechanical flexibility, processability, water solubility, low thermal conductivity. Its electrical conductivity can be enhanced by treatments that remove non-conducting PSS chains [91]. For all these characteristics, PEDOT:PSS is this an excellent ingredient for flexible thermoelectric composites. In several studies, embedding of PEDOT:PSS and TMD nanosheets was carried out by LPE in different solvents (H$_2$O, dimethylsulfoxide, dimethylformamide, ethanol) and vacuum filtration [91,92,93]. In 1T-WS$_2$/PEDOT:PSS composite thin films the electrical conductivity increased monotonically and Seebeck coefficient decreased monotonically with increasing PEDOT:PSS weight concentration, so that an optimized improved power factor of ~0.045 mW m$^{-1}$ K$^{-1}$ was obtained at 50% wt PEDOT:PSS concentration, which is four times higher than that of the pure restacked 1T-WS$_2$ thin films [91]. In 1T-MoS$_2$/PEDOT:PSS composite thin films, the power factor peaked was maximized at 3-5% wt MoS$_2$ concentration, reaching ~0.046-0.049 mW m$^{-1}$ K$^{-1}$ [92,93]. In these systems, the a very low thermal conductivity ~0.27 W m$^{-1}$ K$^{-1}$ allowed to obtain ZT~0-04 at room temperature [92]. In Fig. 4, thermoelectric properties of TMD nanoflake assemblies and TMD composites prepared by liquid phase exfoliation are collected and compared.

## 6. Tunability of thermoelectric properties in TMDs by electric and magnetic fields

Thanks to their 2D nature, transport properties of TMDs are highly tunable by electric field effect. Extremely high values of the Seebeck coefficient of tens mV/K at room temperature were obtained by driving a single layer MoS$_2$ film into the depletion regime, dominated by variable range hopping transport [94]. More importantly for applications is improving the power factor and field effect can be a powerful tool for reversible and clean optimization of the power factor in TMD samples. In one ML thick MoS$_2$ and WSe$_2$ films, the power factor was tuned by nearly one order of magnitude by ambipolar field effect [95], using ion-gel-double-layer gating, and values in the range 0.2-0.3 mW m$^{-1}$ K$^{-2}$ were reached for p-type WSe$_2$ in the accumulation regime. In a similar experiment of ionic liquid-double-layer gating on two-ML WSe$_2$ flakes the power factor was tuned by one order of magnitude, reaching 4 mW m$^{-1}$ K$^{-2}$ in the p-type accumulation regime and 3 mW m$^{-1}$ K$^{-2}$ in the n-type [96] (data are reported in Fig. 1, upper right box). In other already mentioned works, the power factor of one to few ML thick MoS$_2$ flakes were highly enhanced in the accumulation state [59,60] (data are reported in Fig. 1).

In PtSe$_2$ nanosheets, gap opening and semiconducting regime was obtained with decreasing thickness 70. In this semiconducting regime, transport properties become highly tunable by the field effect gating and extrinsic doping. The largest power factor, around 0.14 mW m$^{-1}$K$^{-2}$, was obtained in a two-ML flake in the accumulation regime for V$_{gate}$=+80V at T=300K.

WSe$_2$ polycrystalline films fabricated by thermally assisted conversion of W films on SiO$_2$/Si substrates at 600 °C were also measured under field effect and the maximum p-type power factor was 1.3x10$^{-4}$ mW m$^{-1}$K$^{-2}$, obtained in the depletion regime [97], and presumably associated with decreased thermal conductivity (not measured).

Regarding the effects of magnetic fields, strong and peculiar dependence of thermoelectric transport on magnetic field is expected in TMDs exhibiting topological properties. This is a timely topic of scientific interest, with some possible perspective applications, for example in ferromagnet/TMD bilayers, where the thermoelectric properties of the TMD are enhanced by the static field provided by the ferromagnet.

WTe$_2$ is a type-II Weyl semimetal, whose magnetoelectric transport properties are characterized by large mobilities up to few thousands cm$^2$/Vs at 2K and by non-saturating magnetoresistance related to charge compensation. Thermoelectric properties of μm-sized flakes showed oscillating Seebeck coefficient, due to filling of electron and hole Landau levels [98]. In the same sample, the Nernst effect was observed to be very large at low temperature and non-linear in field due to Weyl points in the energy dispersion relationship. Moreover, in WTe$_2$ valley and spin degrees of freedom are coupled. In ref. [99], the spin (and valley) degeneracy was lifted by depositing a ferromagnetic NiFe film on top of WSe$_2$, thus creating population imbalance between the two valleys. In this system, a valley Nernst effect was detected, and discriminated from anomalous Nernst effect in NiFe and ordinary Nernst effect in both WSe$_2$ and NiFe, resulting in a value $\nu_{VNE} \approx 39$ pA/K.

**7. Devices**
A practical fabrication of an all TMD-based flexible thermoelectric generator was realized, using chemically exfoliated WS$_2$ and NbSe$_2$ nanosheet assemblies [100]. These TMDs were selected for their fairly good power factors, namely n-type power factor $\approx 0.005$-$0.007$ mW m$^{-1}$K$^{-2}$ in WS$_2$ and p-type power factor $\approx 0.026$-$0.034$ mW m$^{-1}$K$^{-2}$ in NbSe$_2$, fairly good values for nanosheet assemblies. Steady operation with up to 38 nW output power at $\Delta T=60$ K was demonstrated, with the benefit of stable performance after 100 bending cycles and after 100 stretching cycles.

A proposal for a TMD-based solid-state thermionic device, operating as power generator or refrigerator, was also presented [101]. The design was based on a van der Waals heterostructure with TMDs (WSe$_2$ and MoSe$_2$) sandwiched between two graphene electrodes. By taking the advantage of the ultralow cross-plane thermal conductance of the 2D materials and the thermionic emission over the Schottky barrier contact between the graphene and 2D materials (tunable via gate voltage or chemical doping), the authors claim that it should be possible to achieve high energy conversion efficiency in the temperature range of 400 to 500 K, comparable to that of traditional thermoelectric devices.

**8. Simulation of performance of a micron-sized thermoelectric module based on optimized TMDs**
In order to assess practically the potential of TMDs as thermoelectric components, we simulated the performance of an ideal thermoelectric module based on p-type and n-type TMD legs, which operates as either cryocooler or power generator. Finite element simulation was carried out with the COMSOL Multiphysics software package. Within this package, the "heat transfer" and "electric currents" physics components, plus the "thermoelectric effect" and "electromagnetic Joule heating" multiphysics components, were used. As material parameters of TMD components (electrical conductivity, Seebeck coefficient, thermal conductivity), we chose from literature values, selecting the largest power factors at 300K, namely power factor 8.5 mW m$^{-1}$K$^{-2}$ measured in n-type MoS$_2$ flakes [59], and power factor $=40$ μWcm$^{-1}$K$^{-2}$ measured in a p-type 3-ML WSe$_2$ flake under field effect [96]. A thermal conductivity value $\kappa=35$ Wm$^{-1}$K$^{-1}$, typical of single crystalline samples [18], was assumed. A silica glass substrate with low thermal conductivity $\kappa \approx 1.38$ W K$^{-1}$m$^{-1}$ was considered. As for the geometry, depicted in the inset of Fig. 5, a source of dissipated heat was placed in the middle, surrounded by a 50 μm deep groove for thermal insulation. The module was composed of 20 p-type and n-type pairs of TMD legs connected electrically in series by gold bridges and thermally in parallel, surrounding the heat source and suspended like bridges over the groove. Each leg was 20 μm wide and 160 μm long, while thickness was represented by the variable parameter $t_{TMD}$. The number of TMD legs, their spacing and their widths were chosen with the criterion of maximum, yet practically feasible, miniaturization. As for boundary conditions, the lateral sides at millimetric distance away from the central dissipating element were thermally connected to a heat sink at room temperature. All the other free surfaces were assumed to exchange heat with environment by air convection and conduction, with an effective heat transfer coefficient of 20 Wm$^{-2}$K$^{-1}$ for larger areas and 2000 Wm$^{-2}$K$^{-1}$ for micrometric areas [102]. For any thickness $t_{TMD}$ of the TMD legs, an input power P was calculated such that the central dissipating element reached $\approx 100°$C in absence of current in the thermoelectric module, namely P=21 mW for $t_{TMD}=20$nm, P=30 mW for $t_{TMD}=200$nm, and P=50 mW for $t_{TMD}=2$μm. We note that such values are determined by the different thermal conductance of the TMD elements only. Applied currents of opposite signs yield Peltier effect of opposite sign, whereas Joule heating depends on the magnitude of the applied current. Hence, during operation, the total heat flow was determined by the balance between Peltier and Joule effects. In the case of operation as cryocooler, an optimized current was calculated for any given $t_{TMD}$ value, which maximized cooling due to Peltier effect against heating due to Joule effect. Our simulations indicated that for a thickness as low as $t_{TMD}$=20nm, the thermoelectric module lowered the temperature of the central element by 0.24 K, with an optimal applied current of 40 μA. The cooling power increased with increasing thickness. For $t_{TMD}$=200 nm the temperature drop was 1.43 K, with an optimal current 300-325 μA, and for $t_{TMD}$=2 μm the drop was 3.18K, with an optimal current 1600-1700 μA (the latter situation is represented in the main panel of Fig. 5). The coefficients of performance (COP) of the thermoelectric module were calculated as ratios of heat removed from the central element to power supplied to the module, for P=0 and for the current that maximally lowers the temperature of the dissipating element below room temperature. We found COP=0.365 for $t_{TMD}$=20nm at optimal current = 30 μA, COP=0.456 for $t_{TMD}$=200 nm at optimal current = 240 μA, and COP=0.485 for $t_{TMD}$=2 μm at optimal current = 1250 μA.

In the case of operation as power generator, with zero applied current, the same dissipated power values that heat the central element at ≈100°C generated output voltages 0.64 V for $t_{TMD}$=20nm, 0.70 V for $t_{TMD}$=200nm, and 0.47 V for $t_{TMD}$=2μm. As output power is maximum for load resistance equal to the output resistance of the module, we estimated output powers 0.0035 mW for $t_{TMD}$=20nm, 0.041 mW for $t_{TMD}$=200nm, and 0.188 mW for $t_{TMD}$=2μm, with power gains 0.000167, 0.0014 and 0.0038 for $t_{TMD}$=20nm, $t_{TMD}$=200nm, and $t_{TMD}$=2μm respectively. Table I reports the above simulation results.

Clearly, these values of temperature drop and COP in the cryocooler operation and of output powers and power gains in the power generator operation are tiny and by far smaller than any value of any commercial interest. Although in the previous section it was shown that power factors of TMDs, determined by their electronic properties, are similar to those of commercial thermoelectrics, the low performance of the simulated module is mainly due to the large thermal conductivity of TMDs in single crystalline form [18]. In this respect, many strategies may be adopted to decrease thermal conductivity, such as lattice misfit in heterostructures and alloys [78,79,80,81,82], mixed horizontal and vertical orientation in flake assemblies [69,75,103], introduction of disorder with defects [62] and nanoinclusions [63,64]. For example, if a low thermal conductivity as measured in disordered $MoS_2$ films [103] was assumed, temperature drops up to 6 times larger and COP values larger than unity would be obtained. The challenge is thus combining within the same TMD sample the low κ obtained by these strategies and the large power factors obtained in TMD crystals, through a systematic investigation of thermoelectric properties of TMDs, TMD alloys and TMD-based composites. It is also worth noting that this all-TMD module may benefit of scaling capabilities.

## 9. Summary and conclusions

From the reviewed literature, it emerges that the electric, thermoelectric and thermal transport properties of TMDs vary significantly with composition, stoichiometry, field effect, chemical doping, confinement, disorder, pressure and presence of interfaces. For example S can be enhanced to the mV/K range in the field effect induced depletion regime in $MoS_2$, $WSe_2$ and $PtSe_2$ flakes and films [94,95,96,70] and room temperature κ can be suppressed down to values smaller than 1 $Wm^{-1}K^{-1}$ in $SnSe_2$ polycrystals [66] and disordered MoS2 flakes [62] and even down to 0.04-0.05 $Wm^{-1}K^{-1}$ in $WSe_2$ with random stacking of layers [75] and in $SnSe_2/MoSe_2$ lattice-mismatched heterostructures [78].

Considering combined properties is more realistic in view of thermoelectric applications. Table II summarizes literature results of power factors exceeding 1 mW $m^{-1}K^{-2}$ and ZT exceeding 0.1 measured in TMD flakes and polycrystals, as well as significant literature results of power factors and ZT measured in TMD nanoflake assemblies and composites. The largest room temperature power factors, up to 8.5 mW $m^{-1}K^{-2}$, were measured in electrostatically doped two-ML $MoS_2$ flakes [59,60], but remarkable values were also measured in $SnSe_2$ polycrystals [67] (0.8 mW $m^{-1}K^{-2}$), in undoped [72] and Cu-intercalated [73] $TiS_2$ flakes (0.371 mW $m^{-1}K^{-2}$ and 1.7 mW $m^{-1}K^{-2}$, respectively), in electrostatically doped two-ML $WSe_2$ flakes [96] (4 mW $m^{-1}K^{-2}$) and two-ML $PtSe_2$ flakes (0.14 mW $m^{-1}K^{-2}$). The largest room temperature figures of merit were obtained in $SnSe_2$ polycrystals (ZT=0.15) [67], $TiS_2$ flakes (ZT=0.16) [72], Cu-intercalated $TiS_2$ flakes (ZT=0.14) [73]. As much remarkable values were obtained in heterostructures, such as ZT=0.37 $(SnS)_{1.2}(TiS_2)_2$ [80] and ZT=0.2 in organic/$TiS_2$ superlattices [84], at room temperature. Noteworthy, interesting values of room temperature power factor up to 0.217 $\mu Wcm^{-1}K^{-2}$ were also measured in $TiS_2$ nanosheets assemblies, which could be produced by large scale methods. In all the cases where measurements were carried out also above room temperature, both power factors and figures of merit increased with increasing temperature, with measured values ZT=0.63 at 673K in $SnSe_2$ polycrystals [67], ZT=0.45 at 800K in Cu-intercalated $TiS_2$ flakes [73], ZT=0.14 at 760K in $MoO_2/MoS_2$ [63], ZT=0.18 at 1000K in $VMo_2S_4/MoS_2$ [64], ZT=0.13 at 1000K in $SnS_2$ flakes [65]. This suggest that 600K-1000K could be the optimal range of application of TMDs as thermoelectrics. However, it must be mentioned that the issue of degradation in environmental conditions must be addressed for certain compositions such as $TiS_2$ [86] and $1T-MoS_2$ [85].

Our simulation of the performance of an all-TMD micrometric thermoelectric module, operating as either cryocooler or power generator, indicated that the literature results with the best measured power factors are not performant enough for commercial use, mainly due to the high thermal conductivity typically measured in the single crystals where high power factors are obtained. The future challenge is obtaining low κ and high power factors in the very same TMD samples.

This review of literature results indicate that plenty of possibilities in terms of sample forms and preparation methods could lead to multifold routes to manipulate electronic and thermal transport and achieve performance improvements. Moreover, field effect experiments, where electronic parameters determining the thermoelectric power factor can be spanned over a wide range in a reversible and clean way, indicate that there is room for improvement also in terms of optimization of electronic parameters. Finally, heterostructures and composites prepared by cost-effective and large-scale methods offer a unique opportunity to combine material properties and take advantage of interface effects. Relying on the technological advancement in the preparation of graphene-based composites, this route seems to be particularly promising for the exploitation of thermoelectric properties of TMDs.

**Acknowledgements**

This work was supported by the FLAG-ERA JTC2017 project MELODICA "Revealing the potential of transition metal dichalcogenides for thermoelectric applications through nanostructuring and confinement" (http://www.melodica.spin.cnr.it/).

**Table I.** Results of the simulation. See text for details.

| | Cryocooler operation with central element at T≈ 373.15K | | | |
|---|---|---|---|---|
| $t_{TMD}$ (nm) | P (mW) dissipated in the central element | Optimal current (μA) in the module | T drop (K) in the central element from 373.15K | |
| 20 | 21 | 40 | -0.24 | |
| 200 | 30 | 300-325 | -1.43 | |
| 2000 | 50 | 1600-1700 | -3.18 | |
| | Cryocooler operation with central element at T≈ 273.15K | | | |
| $t_{TMD}$ (nm) | P (mW) dissipated in the central element | Optimal current (μA) in the module | T drop (K) in the central element from 273.15K | COP |
| 20 | 0 | 30 | -0.15 | 0.365 |
| 200 | 0 | 240 | -0.85 | 0.456 |
| 2000 | 0 | 1250 | -1.77 | 0.485 |
| | Power generator operation with central element at T≈ 373.15K | | | |
| $t_{TMD}$ (nm) | P (mW) dissipated in the central element | $V_{out}$ (V) | $P_{out}$ (mW) | P gain |
| 20 | 21 | 0.64 | 0.0035 | $1.64 \times 10^{-4}$ |
| 200 | 30 | 0.70 | 0.041 | $1.4 \times 10^{-3}$ |
| 2000 | 50 | 0.47 | 0.188 | $3.8 \times 10^{-3}$ |

**Table II.** Summary of noteworthy values of thermoelectric properties measured in TMD systems. In this Table, power factors larger than 1 mW m$^{-1}$K$^{-2}$ and ZT values larger than 0.1 are reported for flakes, films and heterostructures, while in the case of nanoflake assemblies prepared from liquid phase exfoliation power factors larger than 0.01 mW m$^{-1}$K$^{-2}$ and ZT values larger than 0.01 are reported

.

| Sample | Power factor (mW m$^{-1}$K$^{-2}$) | ZT | Ref. |
|---|---|---|---|
| **flakes, films and heterostrures** | | | |
| two-ML MoS$_2$ flake under field effect | 8.5 mW m$^{-1}$K$^{-2}$ at 300 K | | 59 |
| two-ML MoS$_2$ flake under field effect | 3 mW m$^{-1}$K$^{-2}$ at 300 K | | 60 |
| two-ML WSe$_2$ flakes under field effect | 4 mW m$^{-1}$ K$^{-2}$ at 300 K | | 96 |
| highly oriented MoS$_2$ polycrystals with MoO$_2$ nanoinclusions | | 0.14 at 760K along the cross-plane direction | 63 |
| highly oriented MoS$_2$ polycrystals with VMo$_2$S$_4$ nanoinclusions | | 0.18 at 1000 K | 64 |
| highly oriented SnSe$_2$ hot pressed polycrystals | | 0.22 and 0.15 for the chlorine doped and undoped SnSe$_2$ at 610 K | 66 |
| highly oriented SnSe$_2$ pressed polycrystals | | 0.63 at 673 K 0.15 and at 300 K | 67 |
| highly oriented SnSe$_2$ polycrystals with Cu intercalation and Ba substitution | 1.2 mW m$^{-1}$K$^{-2}$ nearly constant from 300 to 773 K | 0.67 at 773 K and 0.10 at 300 K | 68 |
| SnS$_2$ nanoflakes | | 0.13 at 300 K | 65 |
| TiS$_2$ crystals | | 0.16 at 300 K | 72 |
| highly oriented TiS$_2$ polycrystals with Cu intercalation | 1.7 mW m$^{-1}$ K$^{-2}$ at 325 K | 0.45 at 800 K | 73 |
| 1T'-MoTe$_2$ under applied pressure | 30 mW m$^{-1}$ K$^{-2}$ at 25 K | | 76 |
| (MS)$_{1+x}$(TiS$_2$)$_2$ (M = Pb, Bi, Sn) superlattices | | 0.28-0.37 at 700 K for all compositions | 79 |
| (SnS)$_{1.2}$(TiS$_2$)$_2$ superlattices | | 0.37 at 300 K | 80 |
| (SnS)$_{1.2}$(TiS$_2$)$_2$ superlattices with Sb, Co and Cu substitutions | | 0.11 at 300 K and 0.40 at 720 K | 81, 82. |
| TiS$_2$-organic superlattices | | 0.2 at 300 K and 0.28 at 373 K. | 84 |
| **nanoflake assemblies and composites prepared from liquid phase exfoliation** | | | |
| Al:[TiS$_2$ nanosheets] assemblies | 0.217 mW m$^{-1}$K$^{-2}$ at 300 K | | 86 |
| 1T-MoS$_2$ nanosheet assemblies decorated with Cu cations | 0.0235 mW m$^{-1}$ K$^{-1}$ at 300 K | | 88 |
| 1T-MoS$_2$ nanosheet assemblies decorated with Ag nanoparticles | ~0.030 mW m$^{-1}$ K$^{-1}$ at 300 K | | 89 |
| rGO-MoS$_2$ and rGO-WS$_2$ composites | ~0.015-0.017 mW m$^{-1}$ K$^{-1}$ at 300 K | 0.022-0.025 at 300 K | 90 |
| 1T-WS$_2$/PEDOT:PSS composites | ~0.045 mW m$^{-1}$ K$^{-1}$ at 300K | | 91 |
| 1T-MoS$_2$/PEDOT:PSS composite | ~0.046-0.049 mW m$^{-1}$ K$^{-1}$ at 300K | 0.04 at 300 K | 92,93 |

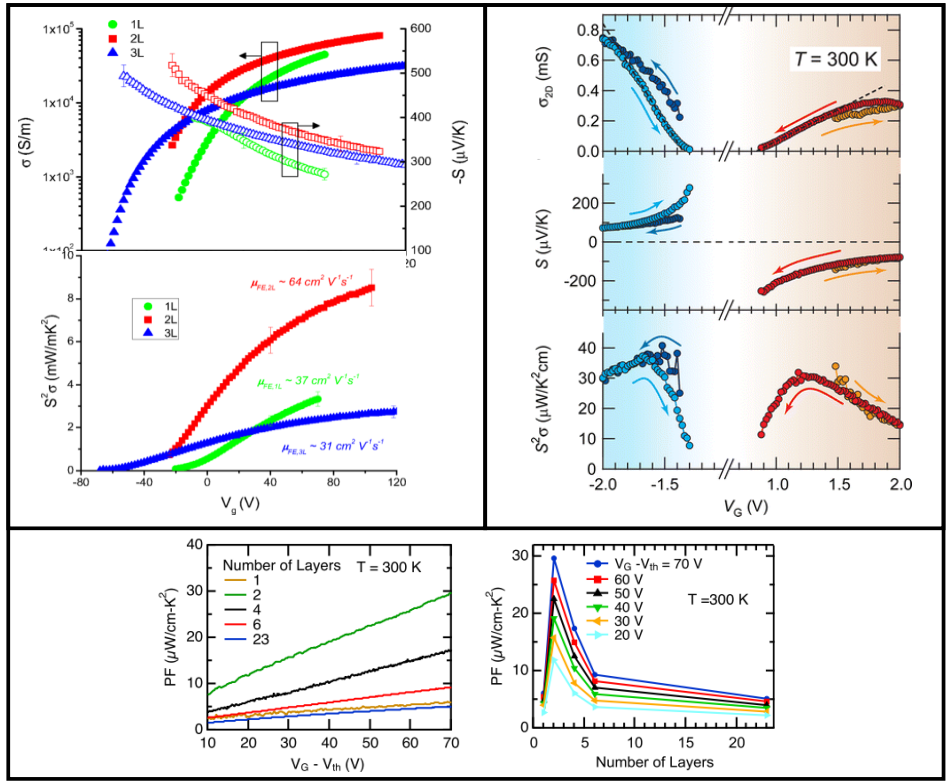

**Figure 1.** Optimized thermoelectric properties of TMD flakes under electric field effect. Upper left box: Electrical conductivities σ, Seebeck coefficients S and power factors S$^2$σ as a function of gate voltage V$_g$ at 300 K for MoS$_2$ monolayer, bilayer and trilayer flakes. Reprinted figure with permission from ref. [59], Copyright (2017) by the American Physical Society. Upper right box: Electrical conductivities σ, Seebeck coefficients S and power factors S$^2$σ as a function of gate voltage V$_g$ at 300 K for a WSe$_2$ bilayer flake. The arrows indicate the direction of gate voltage scan. Reprinted figure with permission from ref. [96]. Copyright (2016) American Chemical Society. Lower bow: Power factors as a function of gate voltage V$_g$ and number of layers measured at 300 K in MoS$_2$ flakes. Reprinted figure with permission from ref. [60], Copyright (2016) by the American Institute of Physics. Note that in these three boxes, the units for power factors are different.

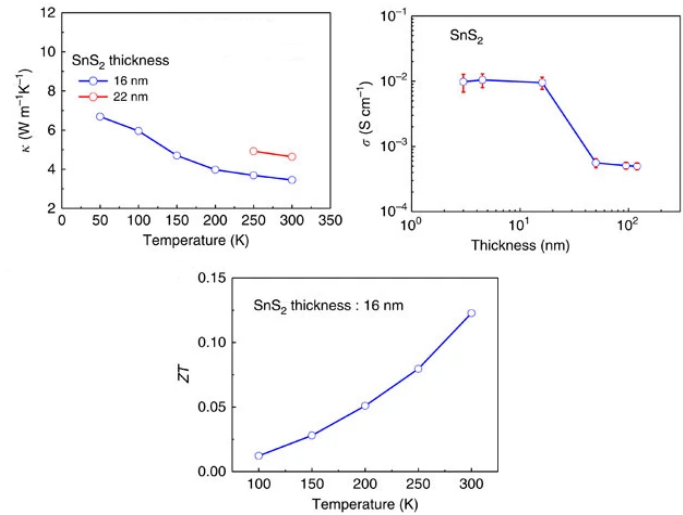

**Figure 2.** Upper left panel: Thermal conductivities $\kappa$ as a function of temperature for 16- and 22-nm-thick SnS$_2$ flakes. Upper right panel: Room temperature electrical conductivities $\sigma$ as a function of SnS$_2$ thickness, showing a large increase below 16 nm and an inverse correlation between $\kappa$ and $\sigma$ with varying thickness. Lower panel: ZT values for temperatures from 100 to 300 K. Reprinted figure from ref. [65].

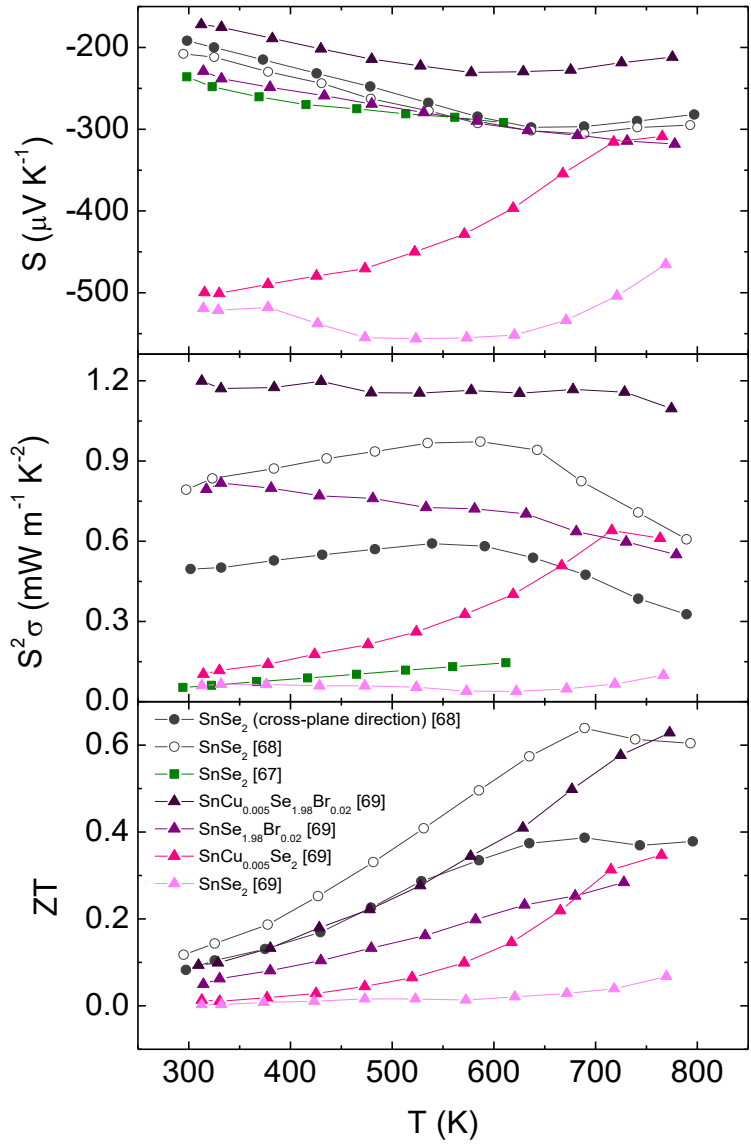

**Figure 3.** Seebeck coefficients S, power factors $S^2\sigma$ and figures of merit ZT as a function of temperature up to 800 K, measured in highly oriented and dense $SnSe_2$ polycrystals prepared by SPS. Data are taken from refs. [66,67,68].

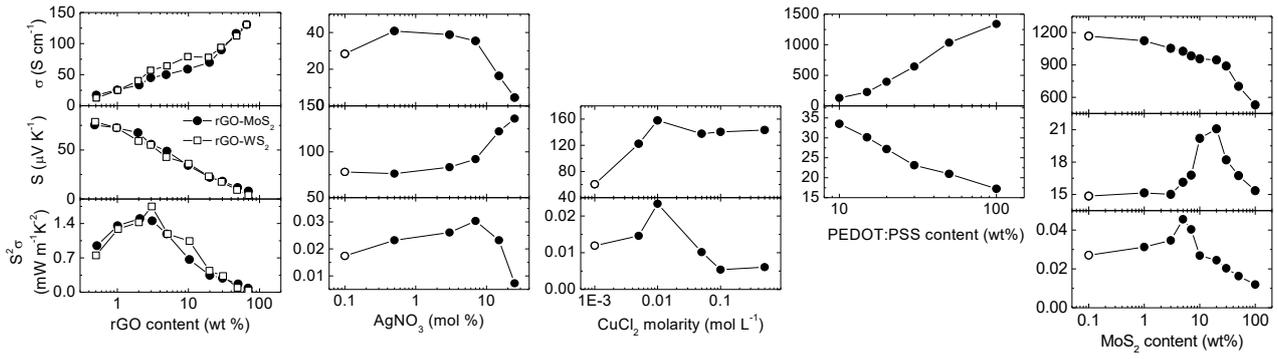

**Figure 4.** Electrical conductivities σ, Seebeck coefficients S and power factors $S^2\sigma$ as a function of component content, measured in TMD-based composites obtained from liquid phase exfoliation of nanoflakes. From left to right: rGO-$MoS_2$ and rGO-$WS_2$ from ref. [90], 1T-$MoS_2$ nanosheets decorated with Ag nanoparticles from ref. [89], 1T-$MoS_2$ nanosheets decorated with Cu cations from ref. [87], 1T-$WS_2$/PEDOT:PSS from ref. [91], 1T-$MoS_2$/PEDOT:PSS from ref. [92]. Note that open round symbols for the lowest concentration in the second, third and fifth panels indicate that the corresponding content value is zero, but was set in this plot to a finite value to be displayed in the logarithmic horizontal scale.

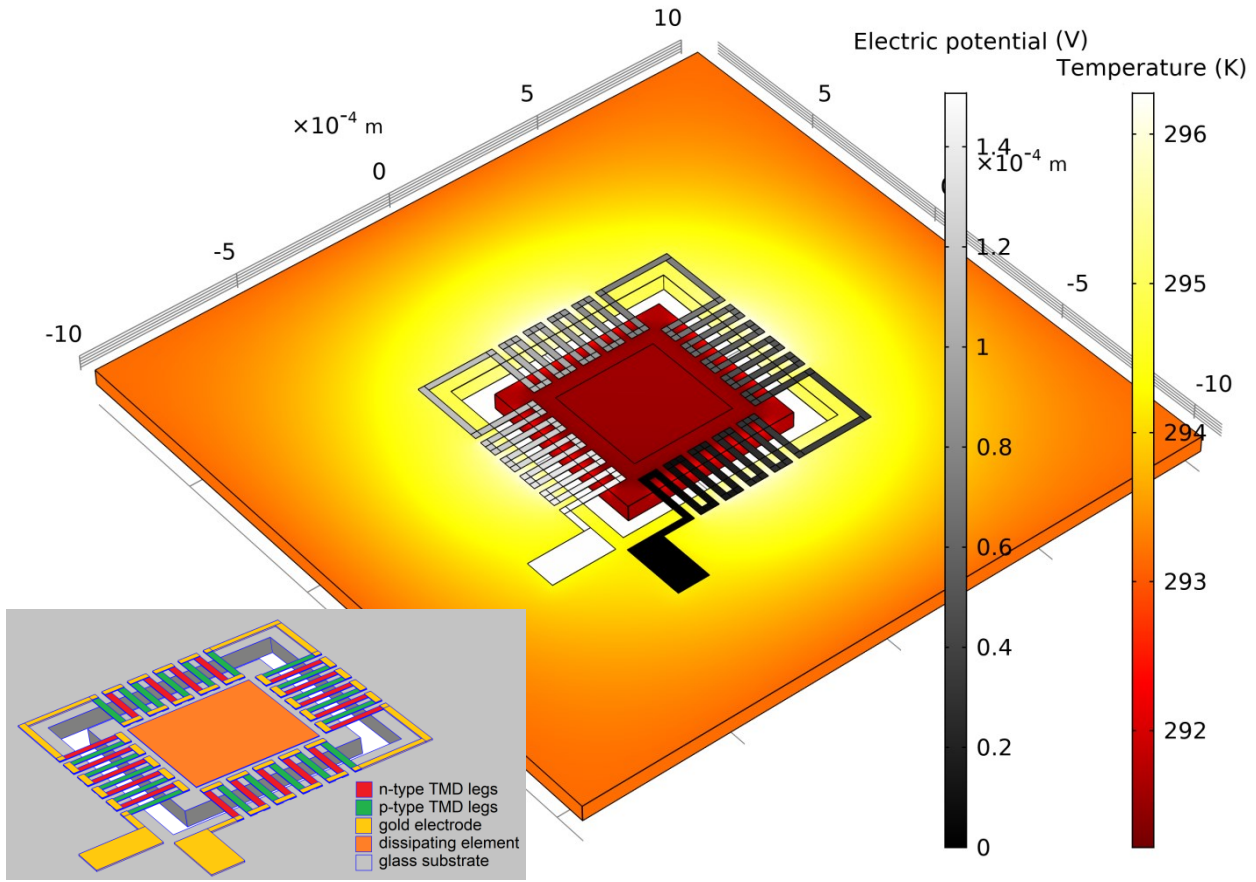

**Figure 5.** Sketch of the micron-sized thermoelectric module, able to operate as either cryocooler or power generator. The overall color scale indicates the temperature, while the color scale in the serpentine of TMD legs indicates the voltage, both calculated as the module is operating as a cryocooler with applied current 1250 µA, which is the optimal value for $t_{TMD}$=2µm and zero dissipated power P=0 in the central element. Inset: material composition of the components of the thermoelectric module.

# 10. References

[1] X. Li, L. Tao, Z. Chen, H. Fang, X. Li, X. Wang, J.-B. Xu, H. Zhu, "Graphene and related two-dimensional materials: Structure-property relationships for electronics and optoelectronics", Applied Physics Reviews 4, 021306 (2017)

[2] S. Manzeli, D. Ovchinnikov, D. Pasquier, O. V. Yazyev, A. Kis, "2D transition metal dichalcogenides", Nature Reviews 2, 17033 (2017)

[3] S.-L. Li, K. Tsukagoshi, E. Orgiu, P. Samorì, "Charge transport and mobility engineering in two-dimensional transition metal chalcogenide semiconductors", Chem. Soc. Rev. 45, 118 (2016)

[4] X. Congxin, L. Jingbo, "Recent advances in optoelectronic properties and applications of two-dimensional metal chalcogenides", Journal of Semiconductors 37(5), 051001 (2016)

[5] H. Hu, A. Zavabeti, H. Quan, W. Zhu, H. Wei, D. Chen, J. Z. Ou, "Recent advances in two-dimensional transition metal dichalcogenides for biological sensing", Biosensors and Bioelectronics 142, 111573 (2019)

[6] K. F. Mak, K. He, J. Shan, T. F. Heinz, "Control of valley polarization in monolayer $MoS_2$ by optical helicity", Nature Nanotechnology 7, 494–498 (2012)

[7] A. A. Tedstone, D. J. Lewis, P. O'Brien, "Synthesis, properties, and applications of transition metal-doped layered transition metal dichalcogenides", Chem. Mater. 28, 1965–1974 (2016)

[8] Y. Y. Hui, X. Liu, W. Jie, N. Y. Chan, J. Hao, Y.-T. Hsu, L.-J. Li, W. Guo, S. P. Lau, "Exceptional tunability of band energy in a compressively strained trilayer $MoS_2$ sheet", ACS Nano, 7 7126-7131 (2013).

[9] H. J. Conley, B. Wang, J. I. Ziegler, R. F. Haglund, S. T. Pantelides, K. I. Bolotin, "Bandgap engineering of strained monolayer and bilayer $MoS_2$", Nano Lett. 13, 3626-3630 (2013)

[10] I. Gutiérrez Lezama, A.. Ubaldini, M. Longobardi, E. Giannini, C. Renner, A. B. Kuzmenko, A. F. Morpurgo, "Surface transport and band gap structure of exfoliated 2H-$MoTe_2$ crystals", 2D Mater. 1, 021002 (2014); I. Gutiérrez Lezama, A. Arora, A. Ubaldini, C. Barreteau, E. Giannini, M. Potemski, and A. F. Morpurgo, "Indirect-to-Direct Band Gap Crossover in Few-Layer $MoTe_2$", Nano Lett. 15, 2336–2342 (2015)

[11] A. Ramasubramaniam, D. Naveh, E. Towe, "Tunable band gaps in bilayer transition-metal dichalcogenides", Phys. Rev. B 84, 205325 (2011)

[12] B. S. Kim, W. S. Kyung, J. J. Seo, J. Y. Kwon, J. D. Denlinger, C. Kim and S. R. Park, "Possible electric field induced indirect to direct band gap transition in $MoSe_2$", Sci. Rep. 7, 5206 (2017)

[13] W. Choi, N. Choudhary, G. H. Han, J. Park, D. Akinwande, Y. H. Lee, "Recent development of two-dimensional transition metal dichalcogenides and their applications", Materials Today 20(3), 116 (2017)

[14] S. Ahmed, J. Yi, "Two-dimensional transition metal dichalcogenides and their charge carrier mobilities in field-effect transistors", Nano-Micro Lett. 9, 50 (2017)

[15] A. Avsar, H. Ochoa, F. Guinea, B. Ozyilmaz, B. J. van Wees, I. J. Vera-Marun," Colloquium: Spintronics in graphene and other two-dimensional materials", Rev. Mod. Phys. (2020) arXiv: 1909.09188

[16] Y.-T. Li, Y. Tian, M.-X. Sun, T. Tu, Z.-Y. Ju, G.-Y. Gou, Y.-F. Zhao, Z.-Y. Yan, F. Wu, D. Xie, H. Tian, Y. Yang, T.-L. Ren, "Graphene-based devices for thermal energy conversion and utilization", Adv. Funct. Mater., 1903888 (2019)

[17] S. Ghosh, W. Bao, D. L. Nika, S. Subrina, E. P. Pokatilov, C. N. Lau, A. A. Balandin, "Dimensional crossover of thermal transport in few-layer graphene" Nat. Mater. 9, 555 (2010)

[18] X. Xu, J. Chen, B. Li, "Phonon thermal conduction in novel 2D materials", J. Phys.: Condens. Matter 28, 483001 (2016)

[19] G. Hu, J. Kang, L. W. T. Ng, X. Zhu, R. C. T. Howe, C. G. Jones, M. C. Hersam, T. Hasan, "Functional inks and printing of two-dimensional materials", Chem. Soc. Rev. 47, 3265 (2018)

[20] X. Gu, Y. Wei, X. Yin, B. Li, R. Yang, "Colloquium : Phononic thermal properties of two-dimensional materials", Rev. Mod. Phys. 90, 041002 (2018)

[21] Y. Ouyang, J. Guo, "A theoretical study on thermoelectric properties of graphene nanoribbons", Appl. Phys. Lett. 94, 263107 (2009)

[22] J. Haskins, A. Kınacı, C. Sevik, H. Sevinçli, G. Cuniberti, T. Çağın, "Control of thermal and electronic transport in defect-engineered graphene nanoribbons", ACS Nano, 5(5), 3779-3787 (2011)

[23] H. Sevinçli, G. Cuniberti, "Enhanced thermoelectric figure of merit in edge-disordered zigzag graphene nanoribbons", Phys. Rev. B 81, 113401 (2010)

[24] Q.-Y. Li, T. Feng, W. Okita, Y. Komori, H. Suzuki, T. Kato, T. Kaneko, T. Ikuta, X. Ruan, K. Takahashi, "Enhanced thermoelectric performance of as-grown suspended graphene nanoribbons", ACS Nano 13(8), 9182-9189 (2019)

[25] F. Ghahari, H.-Y. Xie, T. Taniguchi, K. Watanabe, M. S. Foster, P. Kim, "Enhanced thermoelectric power in graphene: violation of the mott relation by inelastic scattering", Phys. Rev. Lett. 116, 136802 (2016)

[26] J. G. Checkelsky, N. P. Ong, "Thermopower and Nernst effect in graphene in a magnetic field", Phys. Rev. B 80, 081413(R) (2009)

[27] Y. M. Zuev, W. Chang, P. Kim, "Thermoelectric and magnetothermoelectric transport measurements of graphene", Phys. Rev. Lett. 102, 096807 (2009)

[28] P. Wei, W. Bao, Y. Pu, C. N. Lau, J. Shi, "Anomalous thermoelectric transport of Dirac particles in graphene", Phys. Rev. Lett. 102, 166808 (2009)

[29] T. Juntunen, H. Jussila, M. Ruoho, S. Liu, G. Hu, T. Albrow-Owen, L. W. T. Ng, R. C. T. Howe, T. Hasan, Z. Sun, I. Tittonen, "Inkjet printed large-area flexible few-layer graphene thermoelectrics", Adv. Funct. Mater. 1800480 (2018)

[30] L.-D. Zhao, S.-H. Lo, Y. Zhang, H. Sun, G. Tan, C. Uher, C. Wolverton, V. P. Dravid, M. G. Kanatzidis, "Ultralow thermal conductivity and high thermoelectric figure of merit in SnSe crystals", Nature 508, 373-377 (2014)

[31] L.-D. Zhao, G. Tan, S. Hao, J. He, Y. Pei, H. Chi, H. Wang, S. Gong, H. Xu, V. P. Dravid, C. Uher, G. J. Snyder, C. Wolverton, M. G. Kanatzidis, L.D. Zhao, "Ultrahigh power factor and thermoelectric performance in hole-doped single-crystal SnSe", Science 351, 141 (2016)

[32] S. Lee, K. Esfarjani, T. Luo, J. Zhou, Z. Tian, G. Chen, "Resonant bonding leads to low lattice thermal conductivity", Nat. Commun. 5, 3525 (2014)

[33] V. Tayari, B. V. Senkovskiy, D. Rybkovskiy, N. Ehlen, A. Fedorov, C.-Y. Chen, J. Avila, M. Asensio, A. Perucchi, P. di Pietro, L. Yashina, I. Fakih, N. Hemsworth, M. Petrescu, G. Gervais, A. Grüneis, T. Szkopek, "Quasi-two-dimensional thermoelectricity in SnSe", Phys. Rev. B 97, 045424 (2018)

[34] S. Sassi, C. Candolfi, J.-B. Vaney, V. Ohorodniichuk, P. Masschelein, A. Dauscher, B. Lenoir, "Assessment of the thermoelectric performance of polycrystalline p-type SnSe", Appl. Phys. Lett. 104, 212105 (2014)

[35] S. Wang, S. Hui, K. Peng, T. P. Bailey, W. Liu, Y. Yan, X. Zhou, X. Tang, C. Uher, "Low temperature thermoelectric properties of p-type doped single-crystalline SnSe", Appl. Phys. Lett. 112, 142102 (2018)

[36] Q. Zhang, B. Liao, Y. Lan, K. Lukas, W. Liu, K. Esfarjani, C. Opeil, D. Broido, G. Chen, Z. Ren, "High thermoelectric performance by resonant dopant indium in nanostructured SnTe", Proc. Natl. Acad. Sci. USA 110, 13261 (2013)

[37] Q. Tan, L.-D. Zhao, J.-F. Li, C.-F. Wu, T.-R. Wei, Z.-B. Xing, M. G. Kanatzidis "Thermoelectrics with earth abundant elements: low thermal conductivity and high thermopower in doped SnS", J. Mater. Chem. A 2, 17302 (2014)

[38] C.-C. Lin, R. Lydia, J. H. Yun, H. S. Lee, J. S. Rhyee, "Extremely low lattice thermal conductivity and point defect scattering of phonons in Ag-doped $(SnSe)_{1-x}(SnS)_x$ compounds", Chem. Mater. 29, 5344 (2017)

[39] Q. Zhang, E. K. Chere, J. Sun, F. Cao, K. Dahal, S. Chen, G. Chen, Z. Ren, "Studies on thermoelectric properties of n-type polycrystalline $SnSe_{1-x}S_x$ by Iodine doping", Adv. Energy Mater. 5, 1500360 (2015)

[40] S. Liu, N. Sun, M. Liu, S. Sucharitakul, X. P. A. Gao, "Nanostructured SnSe: Synthesis, doping, and thermoelectric properties", J. Appl. Phys. 123, 115109 (2018)

[41] H. Ju, M. Kim, D. Park, J. Kim, "A strategy for low thermal conductivity and enhanced thermoelectric performance in SnSe: Porous $SnSe_{1-x}S_x$ nanosheets", Chem. Mater. 29(7), 3228-3236 (2017)

[42] M. R. Burton T. Liu, J. McGettrick, S. Mehraban, J. Baker, A. Pockett, T. Watson, O. Fenwick, M. J. Carnie, "Thin film Tin Selenide (SnSe) thermoelectric generators exhibiting ultralow thermal conductivity", Adv. Mater. 30, 1801357 (2018)

[43] J. L. Zhang, C. M. Wang, C. Y. Guo, X. D. Zhu, Y. Zhang, J. Y. Yang, Y. Q. Wang, Z. Qu, L. Pi, Hai-Zhou Lu, and M. L. Tian, "Anomalous thermoelectric effects of $ZrTe_5$ in and beyond the quantum limit", Phys. Rev. Lett. 123, 196602 (2019)

[44] W. Zhang, P. Wang, B. Skinner, R. Bi, V. Kozii, C.-W. Cho, R. Zhong, J. Schneeloch, D. Yu, G. Gu, L. Fu, X. Wu, L. Zhang, "Quantized plateau in the thermoelectric Hall conductivity for Dirac electrons in the extreme quantum limit", arXiv:1904.02157v1

[45] W. Huang, X. Luo, C. K. Gan, S. Y. Quek, G. Liang, "Theoretical study of thermoelectric properties of few-layer $MoS_2$ and $WSe_2$", Phys. Chem. Chem. Phys. 16, 10866-10874 (2014)

[46] A. Arab, Q. Li, "Anisotropic thermoelectric behavior in armchair and zigzag mono- and fewlayer $MoS_2$ in thermoelectric generator applications", Sci. Rep. 5, 13706 (2015)

[47] J. Li, J. Shen, Z. Ma, K. Wu, "Thickness-controlled electronic structure and thermoelectric performance of ultrathin $SnS_2$ nanosheets", Sci. Rep. 7: 8914 (2017)

[48] D. Wickramaratne, F. Zahid, R. K. Lake, "Electronic and thermoelectric properties of few-layer transition metal dichalcogenides", J. Chem. Phys. 140, 124710 (2014)

[49] Y. Ge, W. Wan, Y. Ren, Y. Liu, "Large thermoelectric power factor of high-mobility transition-metal dichalcogenides with 1T'' phase", Phys. Rev. Research 2, 013134 (2020)

[50] R. D'Souza, S. Mukherjee S. Ahmad, "Strain induced large enhancement of thermoelectric figure-of-merit (ZT~2) in transition metal dichalcogenide monolayers $ZrX_2$ (X = S, Se, Te)", J. Appl. Phys. 126, 214302 (2019)

[51] A. Shokri, N. Salami, "Thermoelectric properties in monolayer $MoS_2$ nanoribbons with Rashba spin-orbit interaction", J. Mater. Sci. 54, 467-482 (2019)

[52] R.-Z. Zhang, C.-L. Wan, Y.-F. Wang, K. Koumoto, "Titanium sulphene: two-dimensional confinement of electrons and phonons giving rise to improved thermoelectric performance", Phys. Chem. Chem. Phys. 14, 15641–15644 (2012)

[53] Y. Ding, B. Xiao, G. Tang, J. Hong, "Transport properties and high thermopower of $SnSe_2$: a full ab-initio iInvestigation", J. Phys. Chem. C 121, 1, 225-236 (2017)

[54] T. Jia, J. Carrete, Z. Feng, S. Guo, Y. Zhang, G. K.H. Madsen, "Excellent Thermoelectric Performances of Pressure Synthesized ZnSe$_2$", arXiv:1912.10603 (2019)

[55] T. Jia, Z. Feng, S. Guo, X. Zhang, Y. Zhang, "Screening promising thermoelectric materials in binary chalcogenides through high-throughput computations", ACS Appl. Mater. Interfaces 12(10), 11852-11864 (2020)

[56] Z. Zhang, Y. Xie, Q. Peng, Y. Chen, "A theoretical prediction of super high-performance thermoelectric materials based on MoS$_2$/WS$_2$ hybrid nanoribbons", Sci. Rep. 6: 21639 (2016)

[57] G. Ding, C. Wang, G. Gao, K. Yao, C. Dun, C. Feng, D. Li, G. Zhang, "Engineering of charge carriers via a two-dimensional heterostructure to enhance the thermoelectric figure of merit", Nanoscale 10, 7077 (2018)

[58] M. K. Mohanta, A. Rawat, N. Jena, Dimple, R. Ahammed, A. De Sarkar, "Interfacing boron monophosphide with molybdenum disulphide for an ultrahigh performance in thermoelectrics, 2D excitonic solar cells and nanopiezotronics", ACS Appl. Mater. Interfaces 12(2) 3114-3126 (2020)

[59] K. Hippalgaonkar, Y. Wang, Y. Ye, D. Y. Q., H. Zhu, Y. Wang, J. Moore, S. G. Louie, X. Zhang, "High thermoelectric power factor in two-dimensional crystals of MoS$_2$" Phys. Rev. B 95, 115407 (2017)

[60] M. Kayyalha, J. Maassen, M. Lundstrom, L. Shi, Y. P. Chen "Gate-tunable and thickness-dependent electronic and thermoelectric transport in few-layer MoS$_2$", J. Appl. Phys. 120, 134305 (2016)

[61] R. J. Dolleman, D. Lloyd, M. Lee, J. S. Bunch, H. S. J. van der Zant, P. G. Steeneken, "Transient thermal characterization of suspended monolayer MoS$_2$" Phys. Rev. Materials 2, 114008 (2018)

[62] A. Aiyiti, S. Hu, C. Wang, Q. Xi, Z. Cheng, M. Xia, Y. Ma, J. Wu, J. Guo, Q. Wang, J. Zhou, J. Chen, X. Xu, B. Li, "Thermal conductivity of suspended few-layer MoS$_2$", Nanoscale 10, 2727 (2018)

[63] S. Kong, T. Wu, M. Yuan, Z. Huang, Q.-L. Meng, Q. Jiang, W. Zhuang, P. Jiang, X. Bao, "Dramatically enhanced thermoelectric performance of MoS$_2$ by introducing MoO$_2$ nanoinclusions" J. Mater. Chem. A 5, 2004 (2017)

[64] S. Kong, T. Wu, W. Zhuang, P. Jiang, X. Bao, "Realizing p-type MoS$_2$ with enhanced thermoelectric performance by embedding VMo$_2$S$_4$ nanoinclusions" J. Phys. Chem. B 122(2), 713–720 (2018)

[65] M.-J. Lee, J.-H. Ahn, J. H. Sung, H. Heo, S. Gi Jeon, W. Lee, J. Y. Song, K.-H. Hong, B. Choi, S.-H. Lee, M.-H. Jo, "Thermoelectric materials by using two-dimensional materials with negative correlation between electrical and thermal conductivity", Nat. Commun. 7, 12011 (2016)

[66] S. Saha, A. Banik, Dr. K. Biswas, "Few-layer nanosheets of n-type SnSe$_2$", Chem. Eur. J. 22, 15634-15638 (2016)

[67] Y. Luo, Y. Zheng, Z. Luo, S. Hao, C. Du, Q. Liang, Z. Li, K. A. Khor, K. Hippalgaonkar, J. Xu, Q. Yan, C. Wolverton, M. G. Kanatzidis, "n-type SnSe$_2$ oriented-nanoplate-based pellets for high thermoelectric performance", Adv. Energy Mater. 8, 1702167 (2018)

[68] C. Zhou, Y. Yu, X. Zhang, Y. Cheng, J. Xu, Y. K. Lee, B. Yoo, O. Cojocaru-Mirédin, G. Liu, S.-P. Cho, M. Wuttig, T. Hyeon, I. Chung, "Cu intercalation and Br doping to thermoelectric SnSe$_2$ lead to ultrahigh electron mobility and temperature-independent power factor", Adv. Funct. Mater. 1908405 (2019) doi.org/10.1002/adfm.201908405

[69] J. Chen, D. M. Hamann, D. Choi, N. Poudel, L. Shen, L. Shi, D. C. Johnson, S. Cronin, "Enhanced cross-plane thermoelectric transport of rotationally disordered SnSe$_2$ via Se-vapor annealing", Nano Lett. 18(11), 6876-6881 (2018)

[70] H. Moon, J. Bang, S. Hong, G. Kim, J. W. Roh, J. Kim, W. Lee, "Strong thermopower enhancement and tunable power factor via semimetal to semiconductor transition in a transition-metal dichalcogenide", ACS Nano (2019), DOI: 10.1021/acsnano.9b06523

[71] C. A. Kukkonen, W. J. Kaiser, E. M. Logothetis, B. J. Blumenstock, P. A. Schroeder, S. P. Faile, R. Colella, J. Gambold, "Transport and optical properties of Ti$_{1+x}$S$_2$", Phys. Rev. B 24, 1691 (1981)

[72] H. Imai, Y. Shimakawa, Y. Kubo, "Large thermoelectric power factor in TiS$_2$ crystal with nearly stoichiometric composition", Phys. Rev. B 64, 241104(R) (2001)

[73] E. Guilmeau, Y. Bréard, A. Maignan, "Transport and thermoelectric properties in Copper intercalated TiS$_2$ chalcogenide", Appl. Phys. Lett. 99, 052107 (2011)

[74] D. Li, X.Y. Qin, J. Zhang, H.J. Li, "Enhanced thermoelectric properties of neodymium intercalated compounds Nd$_x$TiS$_2$", Physics Letters A 348, 379–385 (2006)

[75] C. Chiritescu, D. G. Cahill, N. Nguyen, D. Johnson, A. Bodapati, P. Keblinski, P. Zschack, "Ultralow thermal conductivity in disordered, layered WSe$_2$ crystals", Science 315(5810), 351-353 (2006)

[76] H. Takahashi, K. Hasegawa, T. Akiba, H. Sakai, M. S. Bahramy, S. Ishiwata, "Giant enhancement of cryogenic thermopower by polar structural instability in the pressurized semimetal MoTe$_2$", Phys. Rev. B 100, 195130 (2019)

[77] H. Sakai, K. Ikeura, M. S. Bahramy, N. Ogawa, D. Hashizume, J. Fujioka, Y. Tokura, S. Ishiwata, "Critical enhancement of thermopower in a chemically tuned polar semimetal MoTe$_2$" Sci. Adv. 2, e1601378 (2016)

[78] E. Hadland, H. Jang, M. Falmbigl, R. Fischer, D. L. Medlin, D. G. Cahill, D. C. Johnson, "Synthesis, characterization, and ultralow thermal conductivity of a lattice-mismatched SnSe$_2$(MoSe$_2$)$_{1.32}$ heterostructure", Chem. Mater. 31(15), 5699-5705 (2019)

[79] C. Wan, Y. Wang, N. Wang, K. Koumoto, "Low-thermal-conductivity $(MS)_{1+x}(TiS_2)_2$ (M = Pb, Bi, Sn) misfit layer compounds for bulk thermoelectric materials", Materials 3, 2606-2617 (2010)

[80] C.Wan, Y. Wang, N. Wang, W. Norimatsu, M. Kusunoki, K. Koumoto, "Development of novel thermoelectric materials by reduction of lattice thermal conductivity", Sci. Technol. Adv. Mater. 11, 044306 (2010)

[81] C. Yin, H. Liu, Q. Hu, J. Tang, Y. Pei, R. Ang, "Texturization induced in-plane high-performance thermoelectrics and inapplicability of the Debye model to out-of-plane lattice thermal conductivity in misfit-layered chalcogenide", ACS Appl. Mater. Interfaces 11, 48079-48085 (2019)

[82] C. Yin, Q. Hu, M. Tang, H. Liu, Z. Chen, Z. Wang, R. Ang, "Boosting the thermoelectric performance of misfit-layered $(SnS)_{1.2}(TiS_2)_2$ by a Co-and Cu-substitutedalloying effect", ., J. Mater. Chem. A 6, 22909 (2018)

[83] Y. E. Putri, C. Wan, F. Dang, T. Mori, Y. Ozawa, W. Norimatsu, M. Kusunoki, K. Koumoto, "Effects of transition metal substitution on the thermoelectric properties of metallic $(BiS)_{1.2}(TiS_2)_2$ misfit layer sulfide", Journal of Electrochemical Materials 43(6), 1870 (2014)

[84] C. Wan, X. Gu, F. Dang, T. Itoh, Y. Wang, H. Sasaki, M. Kondo, K. Koga, K. Yabuki, G. J. Snyder, R. Yang, K, Koumoto, "Flexible n-type thermoelectric materials by organic intercalation of layered transition metal dichalcogenide $TiS_2$", Nature Materials 14, 622–627 (2015)

[85] H. Huang, Y. Cui, Q. Li, C. Dun, W. Zhou, W. Huang, L. Chen, C. A. Hewitt and D. L. Carroll, "Metallic 1T phase $MoS_2$ nanosheets for high-performance thermoelectric energy harvesting" Nano Energy 26, 172–179 (2016)

[86] Y. Zhou, J. Wan, Q. Li, L. Chen, J. Zhou, H. Wang, D. He, X. Li, Y. Yang, H. Huang, "Chemical welding on semimetallic $TiS_2$ nanosheets for high-performance flexible n-type thermoelectric films" ACS Appl. Mater. Interfaces 9, 42430 (2017)

[87] W. Ding, X. Li, F. Jiang, P. Liu, P. Liu, S. Zhu, G. Zhang, C. Liu, J. Xu "Defect modification engineering on a laminar $MoS_2$ film for optimizing thermoelectric properties", J. Mater. Chem. C 8, 1909 (2020)

[88] Y. Zhang, G. D. Stucky, "Heterostructured approaches to efficient thermoelectric materials", Chem. Mater. 26, 837-848 (2014)

[89] X. Li, T. Wang, F. Jiang, J. Liu, P. Liu, G. Liu, J. Xu, C. Liu, Q. Jiang, "Optimizing thermoelectric performance of $MoS_2$ films by spontaneous noble metal nanoparticles decoration", Journal of Alloys and Compounds 781, 744e750 (2019)

[90] T. Wang, C. Liu, F. Jiang, Z. Xu, X. Wang, X. Li, C. Li, J. Xu, X. Yang, "Solution-processed two-dimensional layered heterostructure thin-film with optimized thermoelectric performance", Phys.Chem.Chem.Phys. 19, 17560 (2017)

[91] T. Wang, C. Liu, X. Wang, X. Li, F. Jiang, C. Li, J. Hou, J. Xu, "Highly enhanced thermoelectric performance of $WS_2$ nanosheets upon embedding PEDOT:PSS", Journal of Polymer Science, part B: Polymer Physics 55, 997-1004 (2017)

[92] X. Li, C. Liu, T. Wang, W. Wang, X. Wang, Q. Jiang, F. Jiang, J. Xu, "Preparation of 2D $MoSe_2$/PEDOT:PSS composite and its thermoelectric properties", Mater. Res. Express 4, 116410 (2017)

[93] F. Jiang, J. Xiong, W. Zhou, C. Liu, L. Wang, F. Zhao, H. Liu, J. Xu, "Use of organic solvent-assisted exfoliated $MoS_2$ for optimizing the thermoelectric performance of flexible PEDOT:PSS thin films", J. Mater. Chem. A 4, 5265 (2016)

[94] J. Wu, H. Schmidt, K. K. Amara, X. Xu, G. Eda, B. Özyilmaz, "Large thermoelectricity via variable range hopping in Chemical Vapor Deposition grown singlelayer $MoS_2$", Nano Lett. 14, 2730 (2015)

[95] J. Pu, K. Kanahashi, N. T. Cuong, C.-H. Chen, L.-J. Li, S. Okada, H. Ohta, T. Takenobu, "Enhanced thermoelectric power in two-dimensional transition metal dichalcogenide monolayers", Phys. Rev. B 94, 014312 (2016)

[96] M. Yoshida, T. Iizuka, Y. Saito, M. Onga, R. Suzuki, Y. Zhang, Y. Iwasa, S. Shimizu, "Gate-optimized thermoelectric power factor in ultrathin $WSe_2$ single crystals", Nanolett. 16, 2061 (2016)

[97] W. Y. Kim, H. J. Kim, T. Hallam, N. McEvoy, R. Gatensby, H. C. Nerl, K. O'Neill, R. Siris, G.-T. Kim, G. S. Duesberg, "Field-dependent electrical and thermal transport in polycrystalline $WSe_2$". Adv. Mater. Interfaces 1701161 (2018)

[98] K. Gaurav Rana, F. K. Dejene, N. Kumar, C. R. Rajamathi, K. Sklarek, C. Felser, S. S. P. Parkin, "Thermopower and unconventional Nernst effect in the predicted type-II Weyl semimetal $WTe_2$", Nano Lett. 18(10), 6591-6596 (2018)

[99] M.-T. Dau, C. Vergnaud, A. Marty, C. Beigné, S. Gambarelli, V. Maurel, T. Journot, B. Hyot, T. Guillet, B. Grévin, H. Okuno, M. Jamet, "The valley Nernst effect in $WSe_2$", Nature Communications 10, 5796 (2019)

[100] J. Y. Oh, J. H. Lee, S. W. Han, S. S. Chae, E. J. Bae, Y. H. Kang, W. J. Choi, S. Y. Cho, J.-O Lee, H. K. Baik, T. I. Lee, "Chemically exfoliated transition metal dichalcogenide nanosheet-based wearable thermoelectric generators", Energy Environ. Sci., 9, 1696 (2016)

[101] S.-J. Liang, B. Liu, W. Hu, K. Zhou, L. K. Ang, "Thermionic energy conversion based on graphene van der waals heterostructures", Sci. Rep. 7, 46211 (2017)

[102] O. Ozsun, B. E. Alaca, A. D. Yalcinkaya, M. Yilmaz, M. Zervas, Y. Leblebici, "On heat transfer at microscale with implications for microactuator design", J. Micromech. Microeng. 19, 045020 (2009)

[103] M. Sledzinska, R. Quey, B. Mortazavi, B. Graczykowski, M. Placidi, D. S. Reig, D. Navarro-Urrios, F. Alzina, L. Colombo, S. Roche, C. M. Sotomayor Torres, "Record low thermal conductivity of polycrystalline $MoS_2$ films: tuning the thermal conductivity by grain orientation", ACS Appl Mater Interfaces 9, 37905 (2017)